\documentclass[journal]{IEEEtran}
\usepackage{cite}

\usepackage[pdftex]{graphicx}
\graphicspath{{./}{./figs/}}
\usepackage{epstopdf}

\usepackage{color,soul}
\usepackage{braket}

\begin{document}

\title{Reducing Spontaneous Emission in Circuit Quantum Electrodynamics by a Combined Readout/Filter Technique}

\author{Nicholas~T.~Bronn, Easwar~Magesan, Nicholas~A.~Masluk, Jerry~M.~Chow, Jay~M.~Gambetta and Matthias~Steffen}

\maketitle

\begin{abstract}
Physical implementations of qubits can be extremely sensitive to environmental coupling, which can result in decoherence. While efforts are made for protection, coupling to the environment is necessary to measure and manipulate the state of the qubit. As such, the goal of having long qubit energy relaxation times is in competition with that of achieving high-fidelity qubit control and measurement. Here we propose a method that integrates filtering techniques for preserving superconducting qubit lifetimes together with the dispersive coupling of the qubit to a microwave resonator for control and measurement. The result is a compact circuit that protects qubits from spontaneous loss to the environment, while also retaining the ability to perform fast, high-fidelity readout. Importantly, we show the device operates in a regime that is attainable with current experimental parameters and provide a specific example for superconducting qubits in circuit quantum electrodynamics.
\end{abstract}

%%%%%%%%%%%%%%%%%%%%%%%%%%%%%%%%%%%%%%%%%%%%%%%%%%%%%%%%%%%%%%%%%%%%%%%%
\section{Introduction}
%%%%%%%%%%%%%%%%%%%%%%%%%%%%%%%%%%%%%%%%%%%%%%%%%%%%%%%%%%%%%%%%%%%%%%%%

\IEEEPARstart{S}{pontaneous} emission of radiation can be a dominant source of energy relaxation for a quantum system coupled to an environment. It is possible to suppress this decay by altering the electromagnetic environment seen by the system (referred to as the Purcell effect) \cite{Purcell1946}. In quantum systems such as superconducting quantum bits (qubits), enhancement of qubit lifetime while maintaining qubit control has previously been achieved by enclosing the qubit in a cavity resonator whose fundamental mode is far detuned from the qubit frequency\cite{Blais2004, Wallraff2004, Houck2008}. While this dispersive coupling of the qubit to the resonator helps to minimize decay channels near the qubit frequency, further suppression of these channels is required as system demands continue to increase for larger computations. Recently, methods for engineering impedance mismatches of the Purcell decay channels of the resonator have helped this cause \cite{Reed2010, Jeffrey2014, Kelly2015}. The intuitive idea behind these methods is to place a filter at the qubit frequency after the resonator output to minimize coupling of modes between the qubit and environment.

Here we provide a simple yet flexible technique that compactly integrates the filtering of the unwanted Purcell loss channels directly into the resonator design. We use the fact that the addition of a capacitor in series with the resonator acts as a notch filter below the resonator frequency, and so the capacitor can be tuned such that the filter lies at the qubit frequency. While in principle this achieves our task, for practical designs that do not require dielectrics, it would require an inconveniently large capacitor for the large qubit-resonator detunings common in dispersive readout in circuit quantum electrodynamics (cQED). To make the scheme more practical, we show via a $Y$-$\Delta$ transformation of the circuit that the filter may instead be realized with a very small capacitor that can effectively be tuned as a stray capacitance between the qubit and the external environment. To demonstrate the practicality of the method we provide a full analysis using parameters of a typical transmon style qubit~\cite{Koch2007} coupled to a resonator. We find that the proposed device extends qubit lifetimes while retaining the ability to perform fast dispersive readout of the system. As a final remark, we note that this combined readout/filter technique minimally alters the device footprint which can aid in implementing a scalable architecture for building larger networks of qubits. 

\begin{figure}[t!]
  \centering
  \includegraphics[width=3.5in]{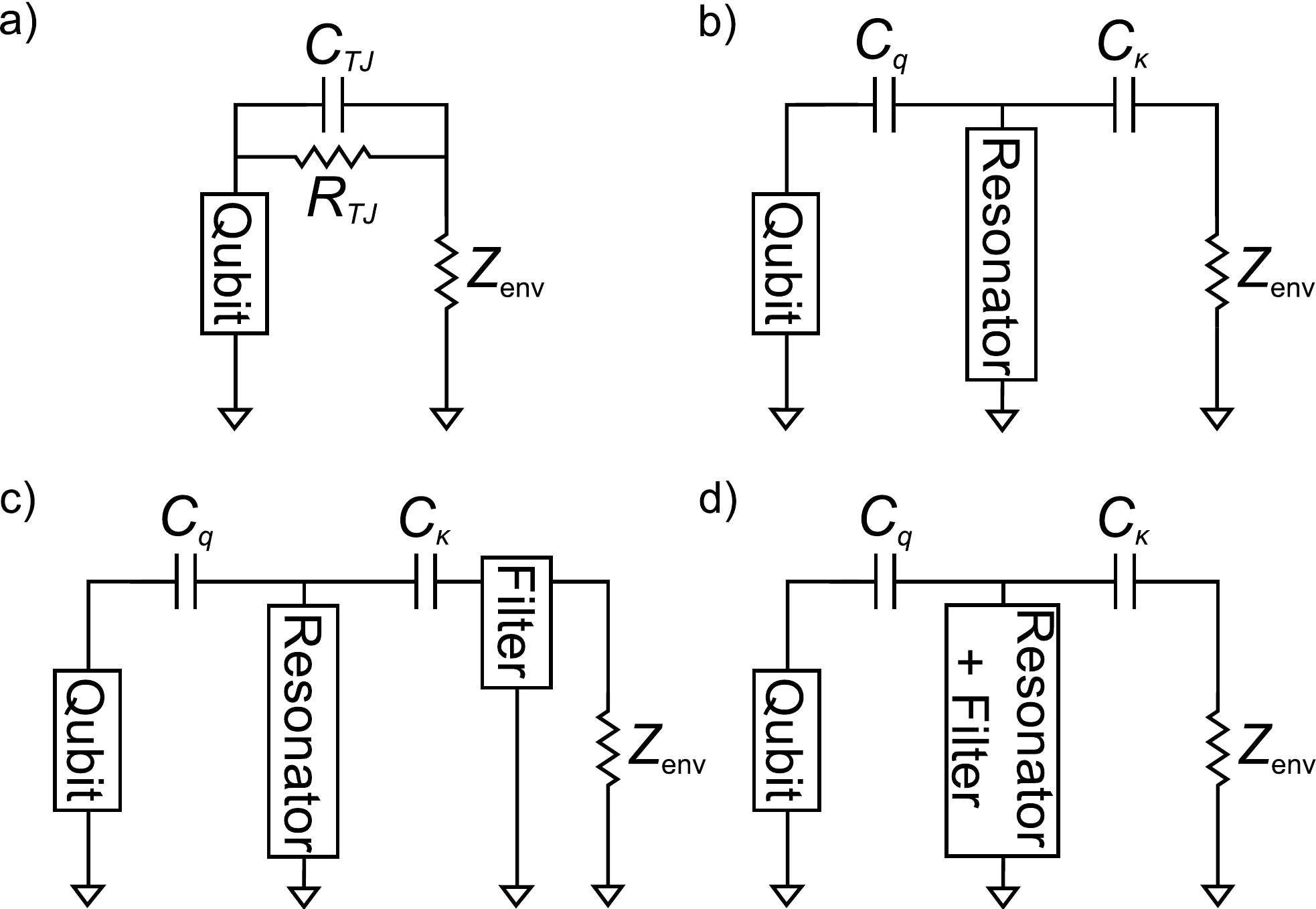}
  \caption{\label{readout-schems} Schematics depicting the evolution of superconducting qubit readout as performed in reflection measurements. a) The state of the original superconducting qubit was read out via tunneling spectroscopy through a tunnel junction of capacitance $C_{TJ}$ and resistance $R_{TJ}$ \cite{Nakamura1997}. b) Spontaneous decay is suppressed by coupling a qubit to the electromagnetic modes of a resonator, as in cQED \cite{Blais2004,Wallraff2004}. $C_q$ is the qubit-resonator coupling capacitor and $C_\kappa$ couples the resonator to the environment (i.e. instruments) c) Recently, researchers have additionally demonstrated filters by implementing impedance mismatches at the qubit decay channels \cite{Reed2010,Jeffrey2014}. d) We propose a simple addition to the readout resonator that allows the same device to simultaneously filter at the qubit frequency.}
\end{figure}

%%%%%%%%%%%%%%%%%%%%%%%%%%%%%%%%%%%%%%%%%%%%%%%%%%%%%%%%%%%%%%%%%%%%%%%%
\section{Background}
%%%%%%%%%%%%%%%%%%%%%%%%%%%%%%%%%%%%%%%%%%%%%%%%%%%%%%%%%%%%%%%%%%%%%%%%
\subsection{Dispersive Readout and Filtering}
%%%%%%%%%%%%%%%%%%%%%%%%%%%%%%%%%%%%%%%%%%%%%%%%%%%%%%%%%%%%%%%%%%%%%%%%

Since the demonstration of coherent control of a solid-state superconducting qubit in 1999~\cite{Nakamura1999}, efforts have been made to protect qubits from dissipation mechanisms. While coupling to the external environment is necessary for the measurement and control of a qubit's state, it is also a major channel for loss. Figure~\ref{readout-schems} depicts the evolution of methods for coupling superconducting qubits to an environment for readout (with $Z_{\rm env} = 50~\Omega$ representing control and measurement instruments). The state of early superconducting qubit devices was read out via the tunnel probe circuit as shown in Fig.~\ref{readout-schems}a, which does not adequately filter the qubit from the $50~\Omega$ environment. Subsequently, leveraging ideas from cavity quantum electrodynamics, a superconducting qubit has also been coupled to a microwave cavity (i.e. coplanar waveguide resonator), launching the field of cQED~\cite{Blais2004,Wallraff2004}. In such a scheme, coupling a qubit to a resonator with discrete electromagnetic modes modifies the available decay channels (see Fig.~\ref{readout-schems}b) from that of the continuum of free space. Depending on the coupling parameters and relevant frequencies, this can greatly enhance or suppress spontaneous emission (the Purcell effect)~\cite{Purcell1946}. Thus, qubit lifetime may be increased in practice by coupling the qubit to a cavity with fundamental frequency $\omega_R$ (and higher-order harmonics) far detuned from the qubit transition frequency $\omega_{\rm ge}$. As the cQED architecture also allows for dispersive control and readout of the qubit state, it has become an attractive option for scaling quantum systems to larger sizes.

Physical implementations of qubits, such as superconducting transmon qubits, are not pure two-level systems, so higher-order transitions must be considered. A transmon coupled to a resonator is described by the Hamiltonian
\[
\hat{H} = \hbar\omega_R \hat{a}^\dagger \hat{a} + \hbar(\omega_{\rm ge} - \frac{\delta}{2})\hat{b}^\dagger \hat{b} + \frac{\hbar\delta}{2} \hat{b}^\dagger \hat{b} \hat{b}^\dagger \hat{b} + \hbar g (\hat{a}\hat{b}^\dagger + \hat{a}^\dagger \hat{b}),
\]
where $\hbar$ is Planck's constant divided by $2\pi$, $\hat{a}$ ($\hat{a}^\dagger$) is the annihilation (creation) operator for photons in the resonator mode $\omega_R$, $\hat{b}$ ($\hat{b}^\dagger$) is annihilation (creation) operator for excitations of the transmon, $\delta$ is the qubit anharmonicity, and $g$ is the qubit-resonator coupling. Although this is not the most standard form of the full, yet simplified, form of Ref. \cite{Koch2007}, this Hamiltonian properly treats the nonlinearity of the transmon caused by its anharmonicity. The familiar Jaynes-Cummings Hamiltonian for a pure two-level qubit can be recovered by letting $\delta \to \infty$. Under the assumption that $g$ is much less than the magnitude of the qubit-resonator detuning, $\Delta = \omega_{\rm ge} - \omega_R$, one obtains the Hamiltonian $\hat{H}_D$ by moving into the dressed basis
\[
\hat{H}_D = \hbar \tilde{\omega}_R \hat{a}^\dagger \hat{a} + \hbar \sum_j \chi_j  \hat{a}^\dagger \hat{a} |j\rangle\langle j| + \hbar \sum_{j\ne 0} \tilde{\omega}_j |j\rangle\langle j|,
\]
where the ket $|j\rangle$ represents the state of the qubit. The parameters $\chi_j$ produce a qubit-state dependent dispersive shift of the resonator frequency $\tilde{\omega}_R$ and a Lamb shift of the qubit frequency $\tilde{\omega}_{\rm ge} = \tilde{\omega}_1$, the first two of which are
\[
\chi_0 = -\frac{g^2}{\Delta} \qquad {\rm and} \qquad \chi_1 = \frac{g^2(\delta-\Delta)}{\Delta(\Delta+\delta)}.
\]
Alternatively, we can rewrite the above Hamiltonian as a photon number-dependent Stark shift of the qubit frequency. When a drive term is added to $\hat{H}$ and rotated into the dressed basis it is straightforward to see that a drive at $\tilde{\omega}_{\rm ge} = \omega_{\rm ge} + \chi_0$ can be used for qubit control, while a drive at $\tilde{\omega}_R = \omega_R \pm \chi$ produces a measurement of the qubit state from the $\chi~=~(\chi_1 - \chi_0)/2$ shift in resonator frequency (or corresponding phase shift). As the measurement drive term commutes with the dressed qubit Hamiltonian the measurement is quantum non-demolition (QND).

The ability to operate large networks of superconducting qubits~\cite{Corcoles2015, Kelly2015} for fault-tolerant error correction protocols, such as the surface code~\cite{Kitaev2003, Bravyi2014, Raussendorf2007}, requires fast, high-fidelity qubit readout. Readout primarily depends on the rate of cavity photon leakage which sets the maximum measurement rate, $\kappa$, the dispersive shift or state-dependent resonator separation, $\chi$, and the strength of the measurement probe, or number of photons $n$. Significantly changing these parameters can be detrimental for a number of reasons, for instance, increasing  $n$ past its critical value of $n_\mathrm{crit} = \Delta^2/4g^2$ can lead to qubit state mixing in the measurement~\cite{Boissonneault2008,Sete2014} and increasing $\chi$ and $\kappa$ enhances cavity-induced Purcell loss \cite{Houck2008}.

Dispersive filtering overcomes the challenge of increasing $\chi$ without affecting $T_1$ by introducing a filter function with a stop-band around the qubit frequency and a pass-band around the resonator frequency. Thus, spontaneous emission of the qubit to environmental modes is greatly suppressed without adversely affecting readout. Current experimentally implemented schemes have been demonstrated where dispersive filtering is achieved with quarter-wave transmission line stubs~\cite{Reed2010} and transmission line bandpass filters~\cite{Jeffrey2014,Kelly2015}. A schematic for this type of configuration is found in Fig.~\ref{readout-schems}c. In this work, we propose the circuit shown in Fig.~\ref{readout-schems}d, and show that a simple modification to the readout resonator conveniently combines the readout function and dispersive filter into one subcircuit. In particular, implementing this design with a quasi-lumped resonator could offer substantial savings in substrate area over separate transmission line resonators and filters.

%%%%%%%%%%%%%%%%%%%%%%%%%%%%%%%%%%%%%%%%%%%%%%%%%%%%%%%%%%%%%%%%%%%%%%%%
\subsection{Relaxation Time}
\label{sec:relax-time}
%%%%%%%%%%%%%%%%%%%%%%%%%%%%%%%%%%%%%%%%%%%%%%%%%%%%%%%%%%%%%%%%%%%%%%%%

Relaxation time is defined as the time constant of the qubit circuit,
\begin{eqnarray} \label{eq:t1-admittance}
T_1^{\rm Purcell}(\omega) = \frac{C_\Sigma}{\mathrm{Re}[Y_q(\omega)]},
\end{eqnarray}
where $C_\Sigma$ is the total qubit capacitance and $Y_q(\omega)$ is the frequency-dependent admittance seen by the qubit. (In the case of the transmon, $C_\Sigma \approx C_J + C_s$, the sum of the Josephson and shunting capacitances). This time constant gives the spontaneous rate of decay due to the Purcell effect. While other qubit relaxation mechanisms such as dielectric loss and two level systems are typically present in a real system, they lie outside the scope of this discussion \cite{Martinis2005,OConnell2008}. By considering $T_1$ from Eq.~\ref{eq:t1-admittance} as a frequency-dependent quantity, we may design filters with desirable stop- and pass-bands.

Making some approximations appropriate for superconducting circuits, it is common to determine the $T_1$ bound from experimentally accessible parameters. The frequency range of operation is GHz, so that $\omega_{\rm ge}, \omega_R \sim 10^{10}$ Hz. Consider the circuit of Fig.~\ref{readout-schems}b, where the resonator is taken as an inductor $L_R$ and capacitor $C_R$ in parallel so that its on-resonant impedance is $Z_R= \sqrt{L_R/C_R}$. The resonator and total qubit capacitors dominate the coupling capacitors, $C_R, C_\Sigma \gg C_q,C_\kappa$, and all capacitors are on the order of $\sim 10-100$ fF. Following the branch-flux method outlined in Ref. \cite{Devoret1997} to calculate the qubit-resonator coupling and applying the aforementioned approximations yields
\begin{eqnarray} \label{eq:g}
g \approx \frac{C_q}{2}\sqrt{\frac{\omega_{\rm ge} \omega_R}{C_J C_R}}.
\end{eqnarray}
Cavity linewidth,
\begin{eqnarray} \label{eq:kappa}
\kappa \approx \omega_{\rm ge}^2 \omega_R Z_R C_\kappa^2 Z_{\rm env},
\end{eqnarray}
is arrived at by using a technique from Ref. \cite{Masluk2012} and subsequent approximations. Then, calculating from Eq.~\ref{eq:t1-admittance} yields
\begin{eqnarray} \label{eq:purcell-loss-eq}
T_1^{\rm cQED}(\omega_{\rm ge}) \approx \frac{\omega_R(\omega_R^2-\omega_{\rm ge}^2)^2}{4\kappa g^2 \omega_{\rm ge}^3},
\end{eqnarray}
and if a small detuning is assumed, we arrive at the familiar approximation $T_1^{-1} \approx \kappa (g/\Delta)^2$. From this expression, it is seen that $T_1$ may be improved by tuning the coupling $g$ to zero, effectively completely isolating the qubit from the environment~\cite{Pashkin2003,Niskanen2007,Allman2010,Gambetta2011}. However, $g$ must be recovered to perform any type of control or measurement via the resonator. Similarly, while a large $\kappa$ is desirable to perform faster readout, increasing $\kappa$ results in a larger resonator-induced spontaneous decay and shorter $T_1$. Eq.~\ref{eq:purcell-loss-eq} does not capture losses due to coupling to higher-order harmonics, which are present in transmission-line resonators, and may also couple to the qubit~\cite{Houck2008}. The presence of these other Purcell decay channels in coplanar waveguide stripline resonators means that it is common to design the qubit transition frequency to be less than that of the resonator fundamental frequency. 

\begin{figure}[t!]
  \centering
  \includegraphics[width=2.5in]{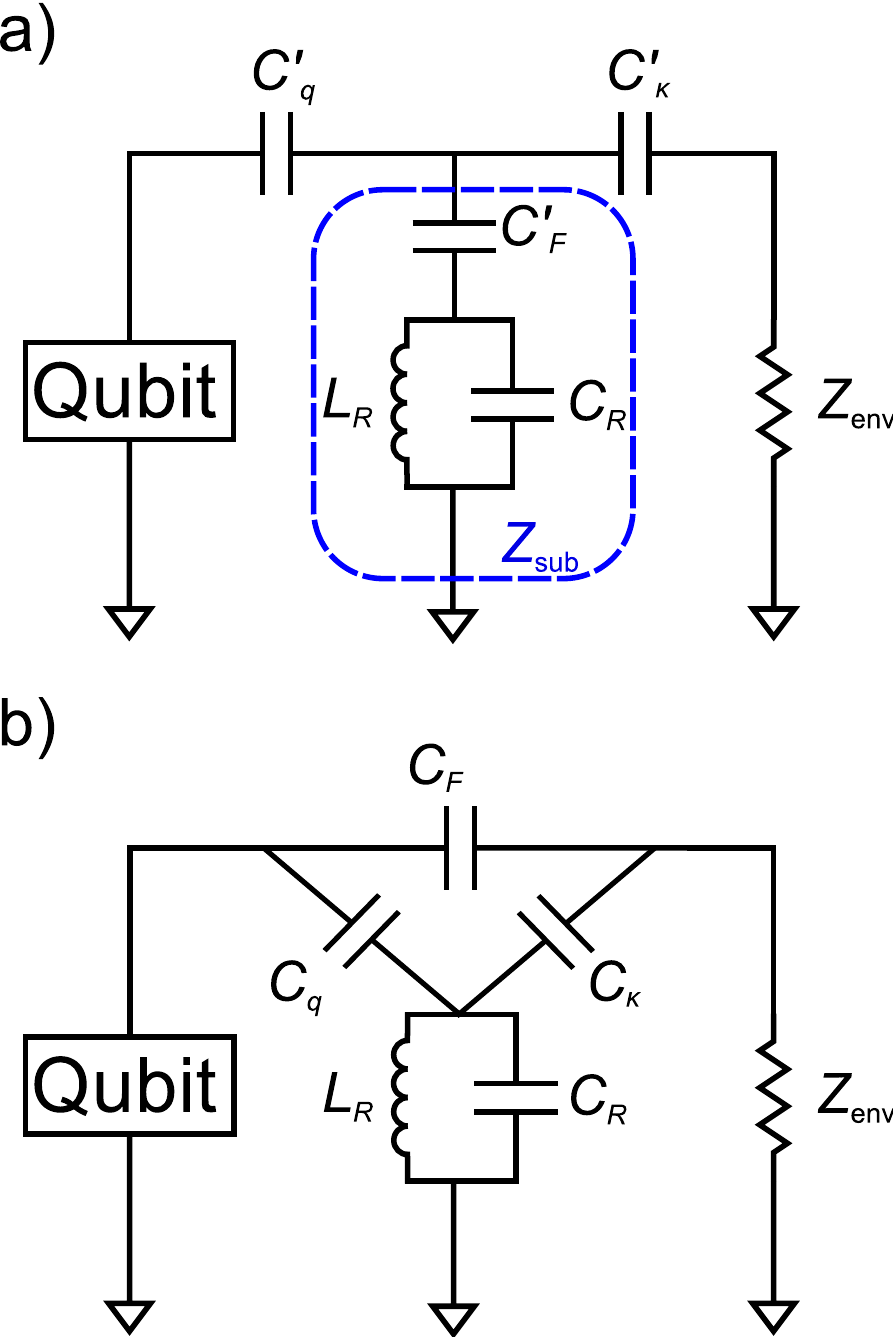}
  \caption{\label{circuit-diags} Circuits for combined readout resonator/notch filter implementation. a) Circuit in the $Y$-configuration is created from the cQED reflection measurement by adding a capacitor $C'_F$ in series with the resonator consisting of $L_R$ and $C_R$. This subcircuit is denoted by an impedance $Z_{\rm sub}$ to ground and used for calculations in the text. b) $\Delta$-configuration equivalent of the circuit in a), which may be a more convenient realization.}
\end{figure}

%%%%%%%%%%%%%%%%%%%%%%%%%%%%%%%%%%%%%%%%%%%%%%%%%%%%%%%%%%%%%%%%%%%%%%%%
\section{Circuit Realization}
%%%%%%%%%%%%%%%%%%%%%%%%%%%%%%%%%%%%%%%%%%%%%%%%%%%%%%%%%%%%%%%%%%%%%%%%
\subsection{Implementation}
%%%%%%%%%%%%%%%%%%%%%%%%%%%%%%%%%%%%%%%%%%%%%%%%%%%%%%%%%%%%%%%%%%%%%%%%

Moving from the cQED readout scheme depicted in Fig.~\ref{readout-schems}b to the combined resonator and filter of Fig.~\ref{readout-schems}d, we again treat the resonator in Fig.~\ref{readout-schems}b as an $LC$ resonator with inductance $L_R$ and capacitance $C_R$ so that $Z_R = \sqrt{L_R/C_R}$ is its on-resonant impedance. Then the combined resonator/filter subcircuit is constructed by placing a filter capacitor $C'_F$ in series with the resonator, as shown in Fig.~\ref{circuit-diags}a. The impedance to ground of this subcircuit is then
\begin{eqnarray*}
Z_{\rm sub}(\omega) & = & \frac{1}{j\omega C'_F} + \frac{1}{j \omega C_R} \bigg|\bigg| j\omega L_R \\
 & = & \frac{1}{j\omega C'_F} + \frac{j\omega L_R}{1-\omega^2L_RC_R} \\
 & = & \frac{1 - \omega^2(L_RC_R + L_RC'_F)}{j\omega C'_F(1-\omega^2/\omega_R^2)}\\
 %& = & \frac{Z_R}{j\omega \omega_R (1/\omega_F^2 - 1/\omega_R^2)}\frac{1 - \omega^2/\omega_F^2}{1-\omega^2/\omega_R^2} \\
 & = & \frac{\omega_R^3 Z_R (\omega_F^2 - \omega^2)}{j\omega (\omega_R^2-\omega_F^2)(\omega_R^2-\omega^2)}
\end{eqnarray*}
where the symbol $||$ indicates the parallel combination of impedances. This addition of the filter capacitor allows $Z_{\rm sub}$ to vanish at the frequency $\omega_F\equiv 1/\sqrt{L_RC_R + L_RC'_F}<\omega_R$. By designing this notch filter to have a frequency commensurate with that of the qubit, $T_1$ is greatly enhanced because the real part of the admittance seen by the qubit also vanishes, as the impedance seen by the qubit is
\[
Z_q(\omega) = \frac{1}{j\omega C'_q} + Z_{\rm sub}(\omega) \bigg|\bigg| \bigg(\frac{1}{j\omega C'_\kappa} + Z_{\rm env} \bigg),
\]
so that $Z_q(\omega_F) = 1/j\omega C'_q$ and the admittance seen by the qubit, $Y_q(\omega_F) = j\omega C'_q$, is purely imaginary. Hence spontaneous emission is greatly suppressed by the combined resonator/filter, without affecting the readout frequency or quality factor of the resonator.  Alternatively, placing a filter inductor in series with the resonator yields a filter above the resonator frequency.

The entire qubit measurement circuit of Fig.~\ref{circuit-diags}a shows the coupling of the combined readout resonator and filter to the qubit by the capacitor $C'_q$ and to the external environment, i.e. measurement and control instruments, by the capacitor $C'_\kappa$. Certain circuit implementations may be more convenient if we perform a $Y$-$\Delta$ transformation of the circuit Fig.~\ref{circuit-diags}a,
\begin{eqnarray*}
C_F & = & \frac{C'_qC'_\kappa}{C'_F+C'_q+C'_\kappa} \\
C_q & = & \frac{C'_\kappa C'_F}{C'_F+C'_q+C'_\kappa} \\
C_\kappa & = & \frac{C'_FC'_q}{C'_F+C'_q+C'_\kappa},
\end{eqnarray*}
into the equivalent circuit of Fig.~\ref{circuit-diags}b, where the unprimed capacitances scale as the inverse of the primed \cite{Kennelly1899}. This analysis may be extended for implementation with other types of planar resonators (i.e. transmission line or stripline) with suitable adjustments.

There is a simple physical explanation for the equivalence between the circuits in Figs.~\ref{circuit-diags}a and b. In the $Y$-configuration of Fig.~\ref{circuit-diags}a, the filter capacitor and resonator form an electrical short to ground at $\omega_F$, thereby acting as a mirror for photons at the qubit frequency, assuming it is commensurate with the notch filter. In the equivalent $\Delta$-configuration of Fig.~\ref{circuit-diags}b, the filter causes interference between photons traveling the two paths from the qubit to the external environment: the path consisting of $C_F$ alone, and the path consisting of $C_q$ and $C_\kappa$. When photons at the qubit frequency are commensurate with the notch filter at $\omega_F$, this interference is destructive, preventing photons from being lost to the environment.

%%%%%%%%%%%%%%%%%%%%%%%%%%%%%%%%%%%%%%%%%%%%%%%%%%%%%%%%%%%%%%%%%%%%%%%%
\subsection{Remaining in the Strong Coupling Regime}
\label{sec:strong-coupling}
%%%%%%%%%%%%%%%%%%%%%%%%%%%%%%%%%%%%%%%%%%%%%%%%%%%%%%%%%%%%%%%%%%%%%%%%

The addition of the filter capacitor must not prevent the ability to control and measure the state of the qubit. As such, we must show that the new device design also allows the qubit and resonator to remain in the strong-dispersive regime of cQED~\cite{Schuster2007}.  Analyzing the configuration in Fig.~\ref{circuit-diags}a for simplicity, we again use the branch-flux method of Ref.~\cite{Devoret1997} to find the Hamiltonian and qubit-resonator coupling. The qubit is modeled as a harmonic oscillator consisting of an inductance $L_J$ and capacitance $C_\Sigma$ in parallel. The coupling is then given by 
\begin{eqnarray} \label{g-eq}
g = \frac{1}{2 \sqrt{Z_1 Z_2}} \frac{\bar{C}_q}{\bar{C}_\Sigma \bar{C}_R + \bar{C}_\Sigma \bar{C_\kappa} + \bar{C}_q \bar{C}_R},
\end{eqnarray}
where $Z_1=\sqrt{L_J/C_1}$ and $Z_2=\sqrt{L_R/C_2}$ are the renormalized on-resonance impedances of the qubit and resonator, respectively, and $C_1$ and $C_2$ are defined by
\begin{eqnarray*}
\frac{1}{C_1} & = & \frac{\bar{C}_R + \bar{C}_q}{ \bar{C}_\Sigma \bar{C}_R + \bar{C}_\Sigma \bar{C_\kappa} + \bar{C}_q \bar{C}_R} \\
\frac{1}{C_2} & = & \frac{\bar{C}_\Sigma + \bar{C}_q}{ \bar{C}_\Sigma \bar{C}_R + \bar{C}_\Sigma \bar{C_\kappa} + \bar{C}_q \bar{C}_R}.
\end{eqnarray*}
The renormalized capacitances $\bar{C}_i$ are related to the circuit parameters of Fig. \ref{circuit-diags}b by
\begin{eqnarray*}
\bar{C}_\Sigma & = & C_\Sigma + C_F \\
\bar{C}_R & = & C_R + C_q \\
\bar{C}_q & = & C_\kappa.
\end{eqnarray*}
Although Eq. \ref{g-eq} is derived using a similar method as the $g$ from Eq. \ref{eq:g}, it is an exact expression. In fact, approximating Eq. \ref{g-eq} yields the same approximate $g$ of Eq. \ref{eq:g}. We consider the effect on qubit-resonator coupling for the specific case of a transmon qubit in Sec.~\ref{sec:transmon}, however first we approximate this expression to estimate qubit lifetime and then analyze how to quantify the overall performance of the measurement process in terms of typical measured parameters. This allows us to parameterize the performance of the measurement in terms of the various system parameters.

%%%%%%%%%%%%%%%%%%%%%%%%%%%%%%%%%%%%%%%%%%%%%%%%%%%%%%%%%%%%%%%%%%%%%%%%
\subsection{Resonator Decay Rate}
\label{sec:res-decay}
%%%%%%%%%%%%%%%%%%%%%%%%%%%%%%%%%%%%%%%%%%%%%%%%%%%%%%%%%%%%%%%%%%%%%%%%

In order to estimate the relaxation time, an expression for resonator linewidth $\kappa$ is needed. Similarly, employing the method of Ref. \cite{Masluk2012} as was used to calculate Eq. \ref{eq:kappa}, the resonator linewidth is
\[
\kappa \approx \frac{\omega_R}{R_{\rm eff} (\omega_R)} \sqrt{\frac{L_R}{C_R + C_{\rm eff}(\omega_R)}},
\]
where
\[
R_{\rm eff}(\omega) = Z_{\rm env} + \frac{(1/C'_\kappa + 1/C'_F)^2}{\omega^2 Z_{\rm env}}
\]
and
\[
C_{\rm eff}(\omega) = \frac{(1/C'_\kappa + 1/C'_F)}{\omega^2 Z_{\rm env}^2 + (1/C'_\kappa + 1/C'_F)^2}.
\]
Under further approximation, this expression yields the same $\kappa$ as that of Eq. \ref{eq:kappa}.

%%%%%%%%%%%%%%%%%%%%%%%%%%%%%%%%%%%%%%%%%%%%%%%%%%%%%%%%%%%%%%%%%%%%%%%%
\subsection{Approximation of Relaxation Time}
\label{sec:approx-t1}
%%%%%%%%%%%%%%%%%%%%%%%%%%%%%%%%%%%%%%%%%%%%%%%%%%%%%%%%%%%%%%%%%%%%%%%%

Here we derive an analytical approximation for the relaxation time as was done in Section \ref{sec:relax-time} for a qubit in the cQED case. This is done to provide the reader with intuition regarding the action of the filter before we return to plotting the exact expressions we derive. As mentioned Sec. \ref{sec:strong-coupling}, the exact qubit-resonator coupling derived in Eq. \ref{g-eq} becomes $g$ of Eq. \ref{eq:g} under our level of approximation. Similarly the approximate $\kappa$ is identical to that in Eq. \ref{eq:kappa}. These approximations will be numerically validated in Secs. \ref{sec:find-g} \& \ref{sec:find-kappa-chi}. The approximate expression for $T_1$ is then
\begin{eqnarray}\label{eq:t1}
T_1^{\rm Filter}(\omega_{\rm ge}) & \approx & \frac{\omega_{\rm ge} (\omega_R^2-\omega_F^2)^2 (\omega_R^2-\omega_{\rm ge}^2)^2}{4\kappa g^2 \omega_R^3 (\omega_F^2 - \omega^2_{\rm ge})^2},
\end{eqnarray}
in which a pole at the filter frequency $\omega_F$ is apparent. Alternatively, qubit lifetime could be expressed as $T_1^{\rm Filter}(\omega)~=~T_1^{\rm cQED}(\omega)/|F(\omega)|^2$, where
\[
F(\omega) = \frac{\omega_R^2 (\omega_F^2-\omega^2)}{\omega^2 (\omega_R^2-\omega_F^2)}
\]
is the filter function acting on the $T_1$ from cQED. Interestingly, this expression as the action of a filter function can be extended to the cQED case, since the resonator acts as a filter at the qubit frequency as well. Consider a qubit coupled to the external environment only through a capacitor $C_\kappa$. Then $T_1^{\rm cap}(\omega)~\approx~C_\Sigma/\omega^2 C_\kappa^2 Z_{\rm env}$, and the action of the cavity as a filter is also apparent as
\[
T_1^{\rm Filter}(\omega) = \frac{T_1^{\rm cQED}(\omega)}{|F(\omega)|^2} = \frac{T_1^{\rm cap}(\omega)}{|F_{\rm cQED}(\omega)|^2|F(\omega)|^2},
\]
where
\[
F_{\rm cQED}(\omega) = \frac{\beta \omega^2}{(\omega_R^2 - \omega^2)}
\]
with $\beta = C_q/C_R$ is the cavity filter function. Here it is seen that the action of multiple filters is multiplicative, as should be the case for filter functions in frequency space.

%%%%%%%%%%%%%%%%%%%%%%%%%%%%%%%%%%%%%%%%%%%%%%%%%%%%%%%%%%%%%%%%%%%%%%%%
\subsection{Measurement Procedure and the Assignment Fidelity}
\label{sec:meas-proc}
%%%%%%%%%%%%%%%%%%%%%%%%%%%%%%%%%%%%%%%%%%%%%%%%%%%%%%%%%%%%%%%%%%%%%%%%

 The measurement signal leaving the resonator, represented by the annihilation operator $b(t)$ of the field mode at time $t$ (we work in the Heisenberg picture), satisfies
 \[
  \left \langle b(t)\right \rangle =\sqrt{\kappa}\left[p_0\alpha_0(t) + p_1 \alpha_1(t)\right]
 \]
\[
= \sqrt{\kappa } \frac{\beta(t)}{2}\langle z \rangle  + \sqrt{\kappa} \frac{\nu(t)}{2}, \label{eq:signal}
\]
 where $\beta(t) =\alpha_0(t)  - \alpha_1(t)$ is the separation between the pointer states $\alpha_0(t)$ and $\alpha_1(t)$ and $\nu(t)/2=(\alpha_0(t)  + \alpha_1(t))/2$ is the mean value of the pointer states. Here $\langle z \rangle = p_0-p_1$ represents the population inversion with $p_0$ and $p_1$ being the probability of the qubit being in state 0 and 1 respectively. 

The mode $b(t)$ undergoes various processing steps before being converted into a measurement result. First, there is an amplification stage which ideally affords the ability to perform single-shot measurements. Under the assumption that the amplifier is narrow bandwidth, linear, and phase-preserving, the output from the amplification stage is given by the field mode $c(t)$
\[
c(t)=\sqrt{G}b(t) + h(t),
\]
where $G$ is the amplifier gain and $h(t)$ is the noise introduced by the amplifier. This signal is then processed via a homodyne (or heterodyne) procedure to obtain full quadrature information and finally the output is digitized and recorded as a trajectory in phase space.

Machine learning classification algorithms can be used to post-process the data and assign measurement outcomes of ``0" or ``1" to the trajectories~\cite{Magesan2014}. Let us suppose the noise affecting the two classes of trajectories is identical, Gaussian-distributed, and has small time correlation. In this case, the standard metric to optimize in machine learning classification is the Fisher separation~\cite{Fisher1936}, defined by
\begin{equation}
{\rm R} =  \frac{(\langle S_0 \rangle-\langle S_1 \rangle )^2}{\mathrm{var}(S_0)}.\label{eq:Fishersep}
\end{equation}
Here $\langle \cdot \rangle$ represents the ensemble mean, $\mathrm{var}( \cdot )$ represents the ensemble variance, and $S_j$ is a real-valued random variable representing the unclassified measurement results for each of $j=0,1$. Loosely speaking $\rm R$ can be thought of as a signal-to-noise ratio as it measures the distance between the means of the ``0" and ``1" classes normalized by the (symmetric) noise in each class. Maximizing $\rm R$ is equivalent to maximizing the assignment fidelity $F$,
\[
F = (1+{\rm erf}(\sqrt{{\rm R}/8}))/2,
\]
 which is a standard metric in quantum information theory for characterizing measurements (here $\mathrm{erf}$ is the error function).

Under these assumptions on the noise, maximizing $\rm R$ (or equivalently $F$) is equivalent to first rotating the quadrature containing the maximal information into the real axis and then performing linear discriminant analysis (LDA), which provides a specific kernel function/classifier (see below) to integrate each signal. The quadrature containing maximal information is defined by $\beta(t) = |\beta(t)|e^{i\theta}$ and so the real-valued quadrature 
\[
I(t)=\mathrm{Re}\left[e^{-i\theta}c(t)\right]
\]
 contains the optimal amount of information in the measurement. The LDA kernel function is given by 
\[
w(t)=\frac{\langle S_0(t) \rangle-\langle S_1(t) \rangle}{\mathrm{var}(S_0(t))} = \frac{\langle I^{(0)}(t)\rangle - \langle I^{(1)}(t)\rangle}{\left|\Delta\left(I^{(0)}(t)\right)\right|^2},
\]
where the superscripts ``(0)'' and ``(1)" denote $|0\rangle$ and $|1\rangle$ state preparations respectively, and $\left|\Delta\left(I^{(0)}(t)\right)\right|^2$ is the variance in $I^{(0)}(t)$. Integrating each real-valued trajectory with $w(t)$ leads to optimal discrimination and maximization of $F$.

Let us now compute the Fisher separation using Ref.~\cite{Magesan2014}. We have
\[
\langle S_0 \rangle =\bigg\langle\int_0^{t_m}w(t)I^{(0)}(t)dt\bigg\rangle = \int_0^{t_m}w(t)\langle I^{(0)}(t)\rangle dt,
\]
which gives
\[
\langle S_0 \rangle =\sqrt{G\kappa}\int_0^{t_m} w(t)\left(e^{-i\theta(t)}\alpha^{(0)}(t) + e^{i\theta(t)}\alpha^{(0)}(t)^*\right)dt.
\]
Similarly for $|1\rangle$ state preparations,
\[
\langle S_1 \rangle =\sqrt{G\kappa}\int_0^{t_m} w(t)\left(e^{-i\theta(t)}\alpha^{(1)}(t) + e^{i\theta(t)}\alpha^{(1)}(t)^*\right)dt,
\]
and so
\[
\left(\langle S_0 \rangle - \langle S_1 \rangle \right)^2 = G\kappa \left(\int_0^{t_m} \frac{|\beta(t)|^2}{\left|\Delta\left(I^{(0)}(t)\right)\right|^2}dt\right)^2.
\]
Since
\[
\left|\Delta\left(I^{(0)}(t)\right)\right|^2 = \frac{G(1+2A)}{4},
\]
where $A \geq \frac{1}{2}\left|1-\frac{1}{G}\right| \geq \frac{1}{2}$ is the added noise of the amplifier normalized by the gain, from Ref. \cite{Magesan2014} we obtain
\begin{equation}
\left(\langle S_0 \rangle-\langle S_1 \rangle \right)^2 = \frac{16\kappa}{G(1+2A)^2}\|\beta(t)\|_2^4.\label{eq:numer}
\end{equation}
Here $\|\beta(t)\|_2 = \sqrt{\int_0^{t_m}|\beta(t)|^2 dt}$ is the two-norm of $\beta(t)$.

Next, $\langle S_0^2 \rangle$ is given by
\[
 \left(\int_0^{t_m} \frac{|\beta(t)|}{\left|\Delta\left(I^{(0)}(t)\right)\right|^2} \left[e^{-i\theta(t)}\langle c^{(0)}(t)\rangle + e^{i\theta(t)}\langle c^{(0)}(t)^\dagger\rangle\right]dt\right)^2
\]
and so
\[
\mathrm{var}(S_0)=\langle S_0^2 \rangle - \langle S_0 \rangle^2 =
\]
\[
 \frac{16}{G^2(1+2A)^2}\left[\int_0^{t_m}\left|\beta(t)\right|^2 \left(\langle e^{-i\theta(t)} c^{(0)}(t)\rangle + e^{i\theta(t)} c^{(0)}(t)^\dagger\rangle \right)dt\right].
\]
Hence
\[
\mathrm{var}(S_0) = \frac{16}{G^2(1+2A)^2}\left[\int_0^{t_m}\left|\beta(t)\right|^2 \left|\Delta\left(I^{(0)}(t)\right)\right|^2\right],
\]
which simplifies to
\begin{equation}
\mathrm{var}(S_0) = \frac{4}{G(1+2A)}\|\beta(t)\|_2^2.\label{eq:denom}
\end{equation}
Combining Eq.'s~\ref{eq:Fishersep},~\ref{eq:numer}, and~\ref{eq:denom} gives
\[
R= \frac{4\kappa}{1+2A}\|\beta(t)\|_2^2.
\]

We can write $R$ completely in terms of physical parameters as follows. The mean number of photons $\bar{n}$ in the resonator is given by
\[
\bar{n} = \frac{2 \mathcal{E}_m^2}{\kappa^2/4 + \chi^2},
\]
where
\[
\chi = \frac{g^2 \delta}{\Delta(\Delta+\delta)}
\]
when higher levels of the transmon are included (transmon anharmonicity given by $\delta$) and $\mathcal{E}_m$ is the measurement drive amplitude. From Ref. \cite{Gambetta2006}, since
\[
||\beta(t)||_2^2 = \frac{4 \chi^2 \mathcal{E}_m^2}{\left(\kappa^2/4 + \chi^2\right)^2},
\]
we obtain
\[
||\beta(t)||_2^2 = \frac{2\chi^2 \bar{n}}{\kappa^2/4 + \chi^2},
\]
and so
\[
R = \frac{8\kappa t_m \chi^2\bar{n}}{(1+2A)\left(\kappa^2/4 + \chi^2\right)}.
\]
Setting $\kappa = 2 \chi$ to maximize $R$ (each photon leaving the resonator carries the maximum amount of information possible about the qubit state) gives
\[
R = \frac{8\chi t_m \bar{n}}{(1+2A)} =  4\kappa t_m \eta \bar{n}
\]
where $\eta = \frac{1}{1+2A}$ is defined to be the efficiency of the amplifier. Hence in total the assignment fidelity is given by
\begin{eqnarray} \label{eq:fidelity-eq}
F = (1+{\rm erf}(\sqrt{{\rm \kappa t_m \eta \bar{n}}/2}))/2.
\end{eqnarray}
This expression for assignment fidelity is valid for infinitely long $T_1$. While it is only valid for a steady state measurement, methods exist the rapidly switch the resonator between off and steady state \cite{McClure2015}.

%%%%%%%%%%%%%%%%%%%%%%%%%%%%%%%%%%%%%%%%%%%%%%%%%%%%%%%%%%%%%%%%%%%%%
\subsection{Application to the Transmon Qubit}\label{sec:transmon}
%%%%%%%%%%%%%%%%%%%%%%%%%%%%%%%%%%%%%%%%%%%%%%%%%%%%%%%%%%%%%%%%%%%%%

Calculations from the preceding sections are now applied to the case of a transmon superconducting qubit with the goal being to show that the new design offers significant improvement in $T_1$ while remaining in the strong-dispersive regime allowing for high-fidelity readout. First let us discuss how we typically choose  numerical values for some of the parameters of our system. This will help give intuition for the relevant regimes when we vary the filter capacitance $C_F$. Afterwards, we will focus on two situations that help show we have achieved our goal. First we analyze how the $T_1(\omega)$ spectrum varies with $C_F$, and find that the $C_F$ can be tuned so that $T_1(\omega)$ has a pole at $\omega_{\rm ge}$. Second, we  show that for the values of $C_F$ that would provide Purcell protection for typical values of $\omega_{\rm ge}$ there is only a small decrease in the transmon-resonator coupling $g$.

%%%%%%%%%%%%%%%%%%%%%%%%%%%%%%%%%%%%%%%%%%%%%%%%%%%%%%%%%%%%%%%%%%%%%%%%
\subsubsection{Typical Numerical Values for Transmon/Resonator Parameters}
%%%%%%%%%%%%%%%%%%%%%%%%%%%%%%%%%%%%%%%%%%%%%%%%%%%%%%%%%%%%%%%%%%%%%%%%

\smallskip

The transmon operates in the regime where the ratio of the Josephson energy $E_J$ over the charging energy $E_C$ is much larger than 1 (typically $E_J/E_C \sim 50-100$). The reason for being in this regime is it allows for significant anharmonicity $\delta \sim -E_C$, necessary to prevent leakage into noncomputational states, while highly suppressing charge noise \cite{Koch2007}. We have $E_C = e^2/(2C_\Sigma)$ where $C_\Sigma = C_J + C_S$ is the sum of the intrinsic Josephson capacitance of the junction and the shunt capacitance, and $E_C$ is typically around $300$~MHz. We choose a qubit frequency of 5~GHz to remain in the transmon regime (minimal charge dispersion), and a resonator frequency of 6.5~GHz to provide a significant dispersive shift (to achieve high-fidelity readout) while preventing hybridization with the qubit. The resonator is modeled as a capacitor $C_R=500$~fF and inductor $L_R=1.2$ nH in parallel. A summary of device parameters discussed here can be found in Table \ref{optimum}.

\begin{table}
  \renewcommand{\arraystretch}{1.3}
  \caption{Summary of Parameters for Combined Filter/Readout Circuit}
  \label{optimum}
  \centering
  \begin{tabular}{|l|c|c|} \hline
    Total qubit capacitance & $C_\Sigma \approx C_J+C_S$ & 65.0 fF \\ \hline
    Qubit inductance & $L_J$ & 15.6 nH \\ \hline
    Qubit frequency & $\omega_{\rm ge}/2\pi$ & 5.0 GHz \\ \hline
    Resonator capacitance & $C_R$ & 500 fF \\ \hline
    Resonator inductance & $L_R$ & 1.2 nH \\ \hline
    Readout frequency & $\omega_R/2\pi$ & 6.5 GHz \\ \hline
    Anharmonicity & $\delta/2\pi$ & -297 MHz \\ \hline
    Environmental Impedance & $Z_{\rm env}$ & 50~$\Omega$ \\ \hline
    Filter capacitance & $C_F$ & 0.50 fF \\ \hline
    Filter capacitance & $C'_F$ & 345 fF \\ \hline
    Filter bandwidth ($T_1 > 1$ ms) & $\Delta \omega_F (1$ ms$) /2\pi$  & 138 MHz \\ \hline
    Filter bandwidth ($T_1 > 10$ ms) & $\Delta \omega_F (10$ ms$) /2\pi$  & 43 MHz \\ \hline
    Qubit coupling capacitance & $C_q$ & 11.1 fF \\ \hline
    Qubit coupling capacitance & $C'_q$ & 12.0 fF \\ \hline
    Qubit-resonator coupling\qquad & $g/2\pi$ & 150 MHz \\ \hline
    Resonator coupling capacitance & $C_\kappa$ & 14.3 fF \\ \hline
    Resonator coupling capacitance & $C'_\kappa$ & 15.4 fF \\ \hline
    Photon decay rate & $\kappa/2\pi$ & 5.0 MHz \\ \hline
    Dispersive shift & $\chi/2\pi$ & 2.5 MHz \\ \hline
  \end{tabular}
\end{table}

%%%%%%%%%%%%%%%%%%%%%%%%%%%%%%%%%%%%%%%%%%%%%%%%%%%%%%%%%%%%%%%%%%%%%%%%
\subsubsection{Selection of Filter Capacitor}
%%%%%%%%%%%%%%%%%%%%%%%%%%%%%%%%%%%%%%%%%%%%%%%%%%%%%%%%%%%%%%%%%%%%%%%%

\smallskip

For the circuit of Fig.~\ref{circuit-diags}a, the selection of filter capacitance with $C'_F \sim 345$~fF would provide enhanced protection for the qubit, however fabrication of such a large capacitor can present challenges. For instance, in our planar architecture, this would require either a large, interdigitated capacitor that produces a parasitic inductance or extra fabrication steps to construct a stacked dielectric capacitor, either of which would likely diminish qubit lifetime. However, considering the $Y$-$\Delta$ transformed circuit of Fig.~\ref{circuit-diags}b, a filter capacitor $C_F~=~0.5$~fF provides the same qubit protection, assuming values for capacitors $C_\kappa$ and $C_q$ that will be determined in Sec.~\ref{sec:find-kappa-chi} to maximize qubit-resonator coupling while remaining in the dispersive regime. The transformed filter capacitance is much smaller and attainable as a stray capacitance between the qubit and external environment. Small-capacitance drive lines on the order of 60 aF have recently been fabricated \cite{Barends2013}. By using relative capacitances based on asymmetrically coupling to our electrically floating qubits, we expect it will be difficult, yet possible, to engineer sub-fF capacitances.

%%%%%%%%%%%%%%%%%%%%%%%%%%%%%%%%%%%%%%%%%%%%%%%%%%%%%%%%%%%%%%%%%%%%%%%%
\subsubsection{$T_1(\omega)$ Spectrum for Various $C_F$ Values}
%%%%%%%%%%%%%%%%%%%%%%%%%%%%%%%%%%%%%%%%%%%%%%%%%%%%%%%%%%%%%%%%%%%%%%%%

\smallskip

\begin{figure}[t!]
  \centering
  \includegraphics[width=3.5in]{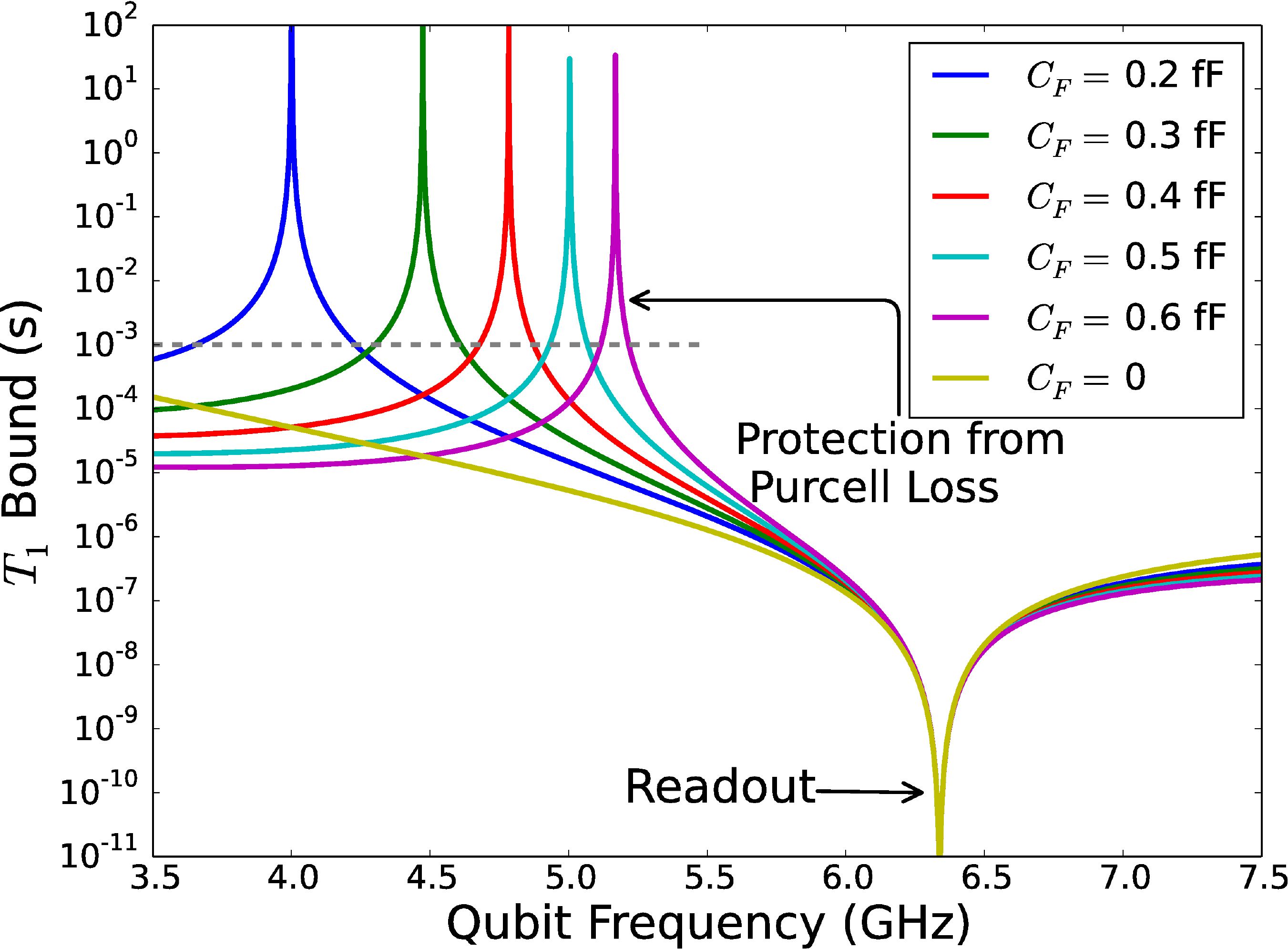}
  \caption{\label{t1} Bound on qubit lifetime, $T_1$, due to the Purcell effect. Protection exhibited for various filter capacitances $C_F$ is much larger than that offered by dispersive readout alone ($C_F=0$ case). The dashed line corresponds to a $T_1$ of a millisecond. The other circuit parameters corresponding to the circuit in Fig.~\ref{circuit-diags}b may be found in Table~\ref{optimum}. The resonator frequency is pulled downward due to the external capacitive coupling of the rest of the circuit.}
\end{figure}

Figure~\ref{t1} shows the maximum $T_1$ bounds as a function of qubit frequency for various non-zero filter capacitances $C_F$, as well as for $C_F=0$ (gold line). The bounds are calculated numerically for the exact expression derived from Eq.~\ref{eq:t1-admittance} and demonstrate the flexibility and enhanced qubit lifetime that the combined readout/filter scheme affords over having no filter. In these calculations, the Josephson inductance $L_J$ is changed while the capacitance $C_J$ remains constant, as this is how the frequency would be tuned experimentally. The qubit-resonator coupling $g$, dispersive shift $\chi$, and cavity decay rate $\kappa$  depend on the qubit and filter frequency through the circuit analysis. However, the approximate forms are independent of $C_F$. For a given qubit frequency $\omega_{\rm ge}$ it is clear that one can tune the capacitance $C_F$ such that $T_1$ is sharply peaked about $\omega_{\rm ge}$. By carefully modeling these circuits with electromagnetic simulations, we can expect to control this capacitance to achieve the desired filtering. In particular, $T_1(\omega)$ spectrum can have bandwidths of hundreds of MHz for lifetime enhancements above a millisecond (see Table~\ref{optimum}). These bandwidths may be increased by reducing $\kappa$ if the fastest measurements are not necessary. 

%%%%%%%%%%%%%%%%%%%%%%%%%%%%%%%%%%%%%%%%%%%%%%%%%%%%%%%%%%%%%%%%%%%%%%%%
\subsubsection{Transmon-Resonator Coupling $g$ for Various $C_F$}
\label{sec:find-g}
%%%%%%%%%%%%%%%%%%%%%%%%%%%%%%%%%%%%%%%%%%%%%%%%%%%%%%%%%%%%%%%%%%%%%%%%

\smallskip

\begin{figure}[t!]
  \centering
  \includegraphics[width=3.5in]{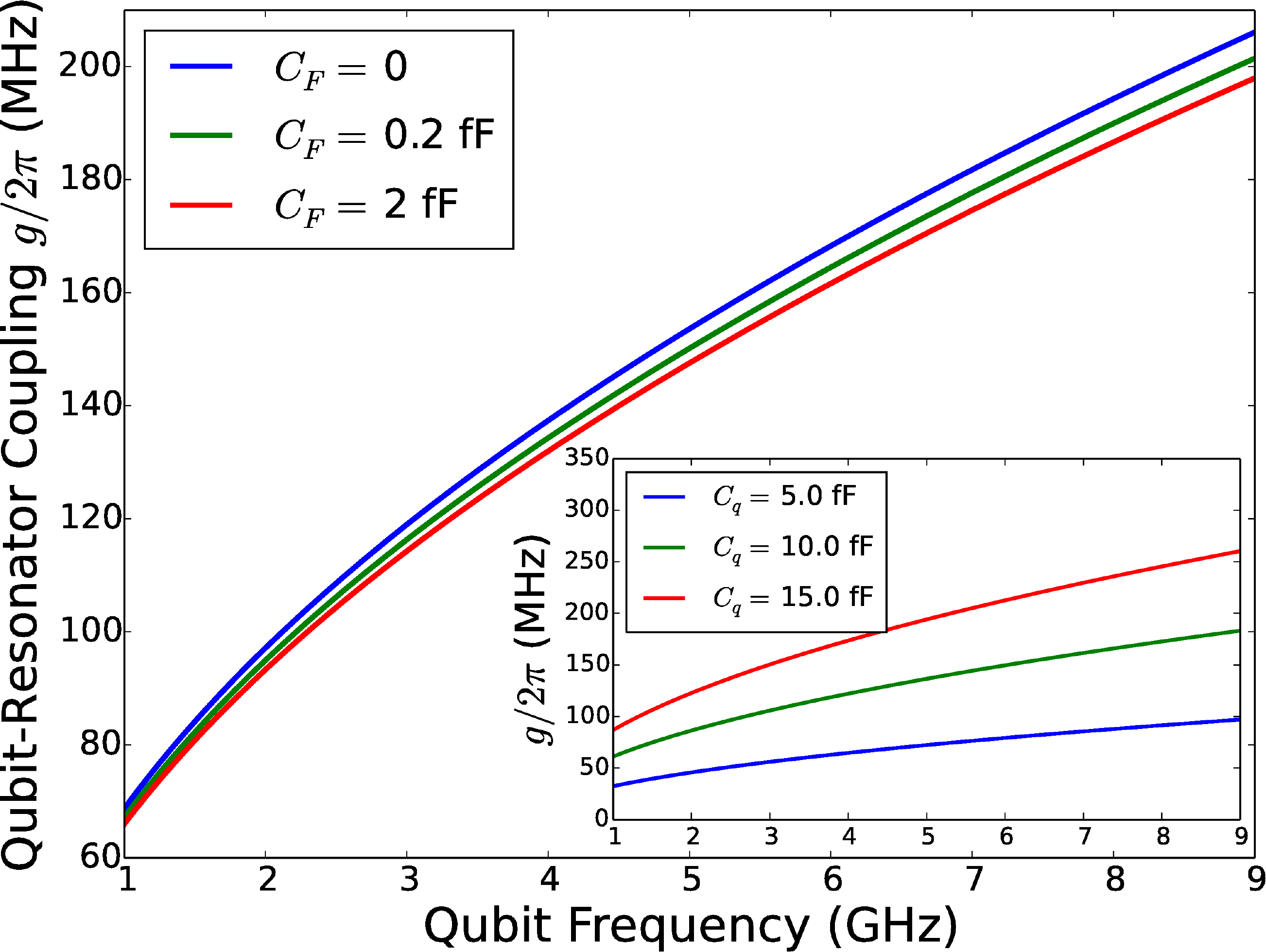}
  \caption{\label{couplings} Qubit-resonator coupling $g$ as a function of qubit frequency for various filter capacitances $C_F$. The inclusion of $C_F$ has a negligible impact on $g$. Inset: qubit-resonator coupling vs qubit frequency for various qubit coupling capacitances $C_q$ showing the large dependence of $g$ with $C_q$.}
\end{figure}

As fast high-fidelity QND measurements require strong coupling between qubit and resonator, the addition of the filter must not significantly degrade the coupling. First, fix the qubit frequency to 5.0~nH by setting the Josephson inductance $L_J = 1.2$~nH. Then Fig.~\ref{couplings}, calculated from Eq.~\ref{g-eq}, demonstrates the negligible effect $C_F$ has on qubit-resonator coupling strength. In particular we find that for any $C_F \in [0,2]$ fF there is small loss in $g$ over the entire spectrum containing typical values of $\omega_{\rm ge}$, validating the approximation for $g$ made in Sec. \ref{sec:approx-t1}. Hence, for a fixed $\omega_{\rm ge}$ and capacitance $C_F$ providing optimal Purcell protection there will be negligible loss in $g$. For the specific case of $C_F = 0.5$ fF, which provide optimal protection, we find the reduction in $g/2\pi$ is only 3.9~MHz. The inset shows that, as expected, the coupling $g$ varies much more strongly with the qubit coupling capacitor $C_q$.

%%%%%%%%%%%%%%%%%%%%%%%%%%%%%%%%%%%%%%%%%%%%%%%%%%%%%%%%%%%%%%%%%%%%%%%%
\subsubsection{Determination of Coupling Capacitors} \label{sec:find-kappa-chi}
%%%%%%%%%%%%%%%%%%%%%%%%%%%%%%%%%%%%%%%%%%%%%%%%%%%%%%%%%%%%%%%%%%%%%%%%

\smallskip

\begin{figure}[t!]
  \centering
  \includegraphics[width=3.5in]{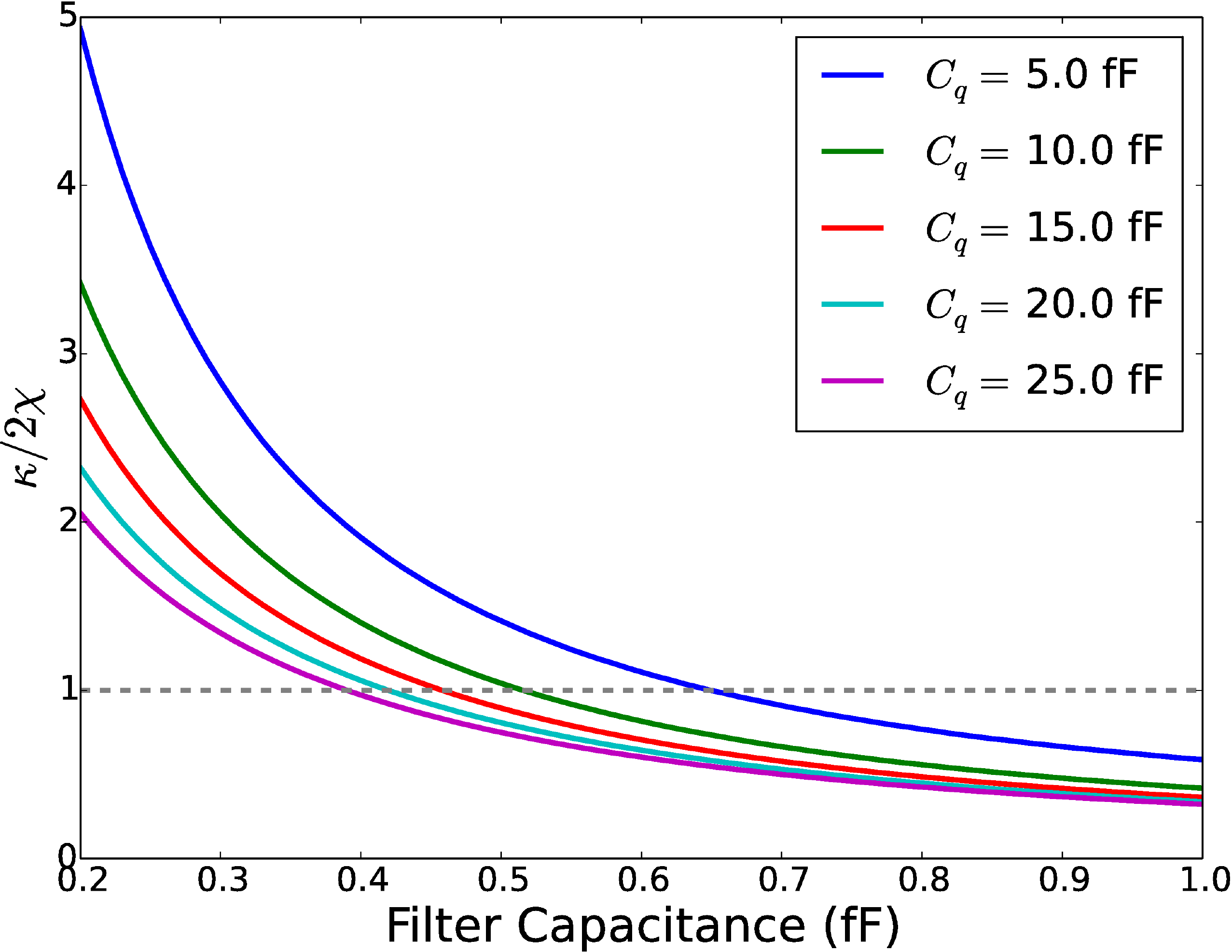}
  \caption{\label{dispersive} $\kappa/2\chi$ as function of filter capacitances $C_F$ for various qubit couplings $C_q$. Signal-to-noise is maximized when $\kappa/2\chi = 1$, indicated by the dashed line~\cite{Gambetta2008}.}
\end{figure}

We now justify the values for coupling capacitors $C_q$ and $C_\kappa$ (or $C'_q$ and $C'_\kappa$) in Table~\ref{optimum} that were used to construct the plots of Figs.~\ref{t1}-\ref{couplings}. These capacitors primarily determine the qubit-resonator coupling $g$ and resonator decay rate $\kappa$, respectively, and therefore are important in determining the quality of the measurement. In general we wish to maximize $\kappa$ in order to perform measurements as quickly as possible, subject to the constraint that Fisher separation is maximized. Optimal Fisher separation occurs when $\kappa = 2\chi$ (from Sec.~\ref{sec:meas-proc}), and Fig.~\ref{dispersive} shows how the ratio $\kappa/2\chi$ depends on filter capacitor and qubit coupling capacitor. In order to remain in the dispersive regime, we must also ensure that $g \ll \Delta$. Choosing $g \sim \Delta/10 = 2\pi \cdot 150$~MHz and applying an iterative method to enforce the other constraints yields coupling capacitances of $C_q = 11.1$~fF and $C_\kappa = 14.3$ fF. These in turn yield a resonator decay rate $\kappa = 5.0$~MHz and dispersive shift $\chi = 2.5$~MHz. While the approximate form of $\kappa$ from Sec. \ref{sec:approx-t1} yields a value of 3.7~MHz, the approximations are still useful for developing intuition about the combined measurement/readout. The full numeric modeling we perform provides the most accurate assessment.

%%%%%%%%%%%%%%%%%%%%%%%%%%%%%%%%%%%%%%%%%%%%%%%%%%%%%%%%%%%%%%%%%%%%%%%%
\subsection{Fast High-Fidelity Measurement of the Transmon} 
%%%%%%%%%%%%%%%%%%%%%%%%%%%%%%%%%%%%%%%%%%%%%%%%%%%%%%%%%%%%%%%%%%%%%%%%

\begin{figure}[t!]
  \centering
  \includegraphics[width=3.5in]{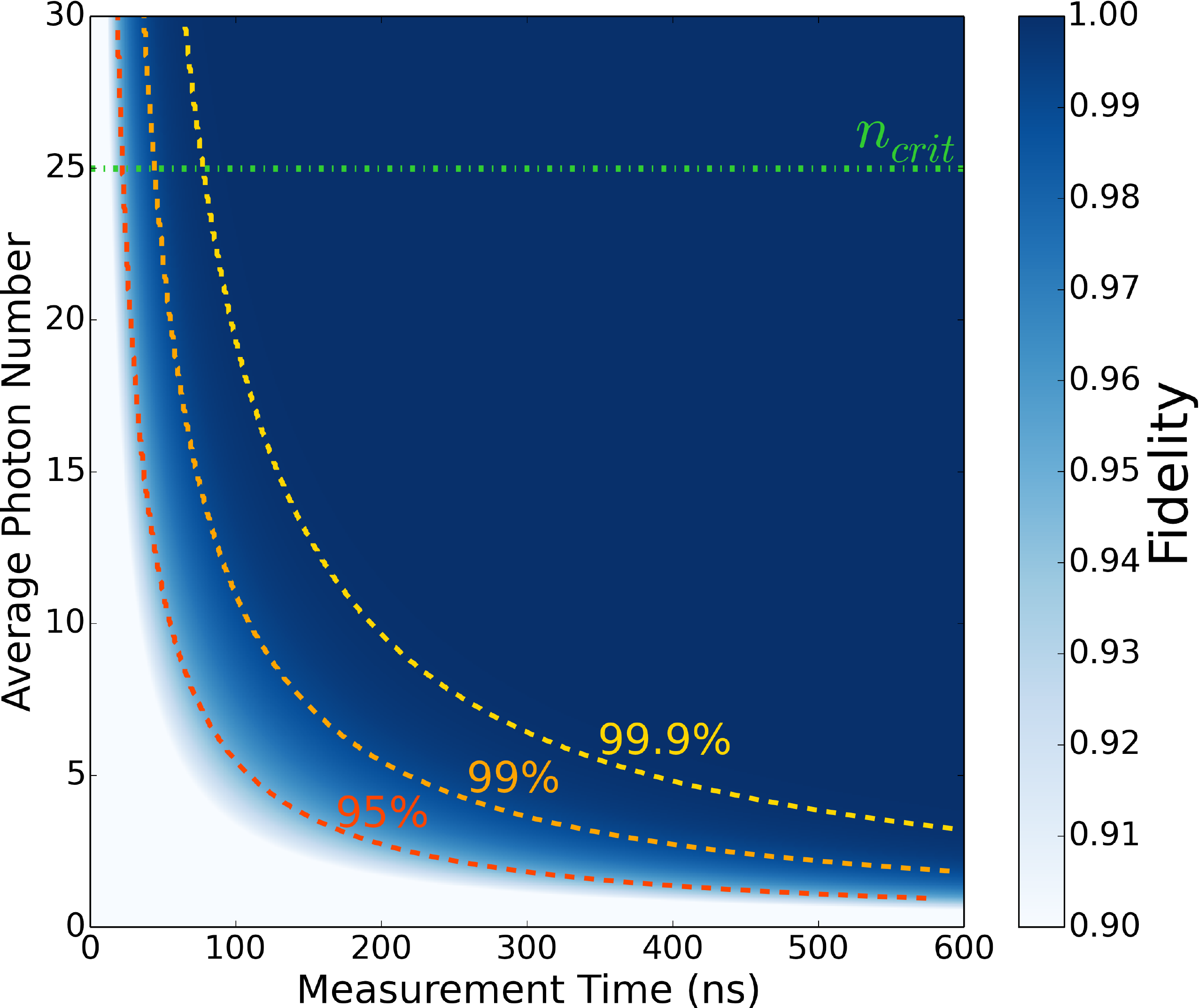}
  \caption{\label{fidelities} A colormap of assignment fidelity given average photon number $\bar{n}$ and measurement time $t_m$ The dashed lines correspond to surfaces of constant fidelity. The dash-dotted line in green is the critical photon number for remaining in the dispersive regime.}
\end{figure}

Fault-tolerant quantum computation requires both fast and high-fidelity measurements. However, the required assignment fidelity, defined in Eq.~\ref{eq:fidelity-eq}, depends on the error model and geometry. If a single photon is used to measure the qubit, a measurement time of 0.55, 1.09, or 1.92~$\mu$s would be required to achieve a fidelity of 95, 99, or 99.9 \%, respectively. Alternatively, assuming the critical number of photons in the resonator ($n_{\rm crit} = 25$) does not affect $\chi$ \cite{Gambetta2006}, the assignment could achieve the respective fidelities with measurement times as little as 22, 44, or 77~ns. New techniques developed to rapidly ring-up and ring-down the resonator may allow these short measurement times and high fidelities to be realized \cite{McClure2015}. However, care must be taken to prevent from Stark shifting the qubit outside of the filter bandwidth. Fig. \ref{fidelities} shows assignment fidelities as a function of average photon number $\bar{n}$ and measurement time $t_m$, with surfaces of constant fidelity highlighted. From this it is clearly seen that fast, high-fidelity measurement may be performed with the combined filter/readout technique.

\section{Conclusion}
We have presented a circuit that combines a qubit readout and dispersive filtering with a slight modification to a readout resonator. By controlling a small stray capacitance between the qubit and the external environment, an anti-resonance appears with the readout resonator that acts as a notch filter. By tuning the notch filter frequency to the qubit transition frequency, spontaneous qubit decay is suppressed. The readout resonator and filter are hence combined, allowing for qubit protection without substantially altering device fabrication or footprint. Thus qubit lifetime is enhanced and the particular case of a superconducting transmon qubit is analyzed. Here the combined filter/readout technique is shown to be compatible with fast, high-fidelity readout in the surface code architecture, holding promise for the advancement of superconducting circuits for quantum computing. 

\section*{Acknowledgment}

This work was supported by IARPA under contract W911NF-10-1-0324. The authors thank Antonio C\'orcoles, Jared Hertzberg, Doug McClure and Will Shanks for insightful discussions. The combined readout/filter contained within is the subject of US Patent Application 14/512489. The authors are with the IBM T.J. Watson Research Center, 1101 Kitchawan Rd, Yorktown Heights, NY, 10598, USA (e-mail: ntbronn@us.ibm.com).
\bibliographystyle{IEEEtran}
\bibliography{paper-purcell}

\end{document}